\documentclass{jpsj3}
\usepackage{txfonts}
\usepackage{color}
\providecommand{\newblock}{}
\title{100-Billion-Atom Molecular Dynamics Simulation of Acoustic Cavitation in a Simple Liquid}

\author{Yuta Asano$^{1, 2}$\thanks{asano.yuta@isu.ac.jp}}
\inst{$^1$Faculty of Global Nursing, Iryo-Sosei University, 1-3-4 Koaota, Kashiwa, Chiba 277-0803, Japan \\
  $^2$Institute for Solid State Physics, The University of Tokyo, Kashiwa, Chiba 277-8581, Japan}

\abst{ A large-scale molecular dynamics (MD) simulation of acoustic cavitation in a simple liquid was performed using the supercomputer Fugaku. The system, consisting of approximately 100 billion atoms, was subjected to ultrasonic irradiation. Direct observation of multi-bubble dynamics has been challenging in both experimental measurements and conventional numerical fluid mechanics simulations. Moreover, previous MD simulations involving only hundreds of millions of atoms were unable to generate multiple bubbles within a system. Our results reveal that cavitation bubbles nucleate and grow near the ultrasonic horn, forming a large bubble cluster that periodically splits into multiple small clusters and subsequently merges again. This cycle is synchronized with the oscillation period of the horn. Pressure and temperature inside the bubbles exhibit sharp increases during cluster fragmentation, and their oscillation amplitudes vary on a timescale longer than the driving period of the horn, indicating the presence of subharmonic behavior consistent with experimental observations. Despite bubble formation, the effect on the acoustic properties of the sound wave was almost negligible, indicating that cavitation near the horn surface has limited influence on bulk acoustic properties. These findings provide new insights into the molecular-scale mechanisms of cavitation and offer guidance for optimizing ultrasonic systems in chemical and biomedical applications. Future work will focus on quantifying long-period oscillations, analyzing attenuation effects, and extending simulations to complex fluids.}
\begin{document}
\maketitle
%\linenumbers
\section{Introduction}
Cavitation is a phenomenon in which bubbles are generated, grow, and collapse in a liquid due to rapid pressure changes~\cite{brennen2014}. It occurs in high-speed flows, such as around rotating screw propellers or inside vortex pumps. When intense ultrasound with a frequency above 20~kHz is applied to a liquid, pressure waves propagate and induce cavitation, known as ultrasonic cavitation. During this process, localized high-temperature and high-pressure regions are formed~\cite{wood1927,richards1927}. Ultrasonic cavitation has been widely applied in medicine, chemistry, and engineering, taking advantage of the extreme physical and chemical environments created during bubble collapse to promote chemical reactions, cleaning, and sterilization~\cite{lajoinie2016,versluis2020,zielewicz2016,xiao2019,adamou2024,hansen2022,boshagh2022}. In particular, its application in drug delivery and non-invasive therapies is expanding, attracting attention from the perspective of safety and environmental compatibility~\cite{lajoinie2016,ferrara2007,roovers2019,wu2008,ma2022}. However, cavitation is an extremely complex phenomenon involving repeated generation and collapse of numerous bubbles, and its mechanism remains insufficiently understood.

When numerous bubbles appear in a cavitation field, experimental studies have shown anomalies such as a decrease in sound speed and an increase in attenuation coefficient~\cite{silberman1957,commander1989}. Furthermore, unlike single-component liquids, sound speed and attenuation exhibit non-monotonic changes with respect to frequency. These anomalies are closely related to bubble dynamics, and theoretical analyses based on the Helmholtz equation have been conducted, with attempts to apply them to sonochemical reactors~\cite{louisnard2012a,louisnard2012b,trujillo2018,garcia2022}. Louisnard's theoretical model has been widely used for parameter searches, enabling qualitative predictions of experimental results. However, several features cannot be described by such mean-field approaches, include the generation of subharmonics caused by multi-bubble dynamics~\cite{neppiras1969}. The mechanism of subharmonic generation is considered to arise from nonlinear bubble oscillations and collective bubble motion. Ultrasonic cavitation often produces bubble streams and characteristic bubble clouds-cylindrical, conical, or mushroom-shaped-through repeated coalescence anfragmentation~\cite{vznidarvcivc2014}. These bubble clouds generate acoustic waves corresponding to their formation-collapse cycle, and the resulting subharmonics are known to enhance chemical reactions. Experimentally, subharmonics tend to occur when the ultrasonic horn diameter and frequency are small~\cite{vznidarvcivc2014}. Nevertheless, the detailed conditions remain unclear due to the complexity of multi-bubble dynamics. Although subharmonic generation has been experimentally observed under specific conditions such as small horn diameter and large amplitude, its underlying mechanism remains elusive. Existing mean-field models cannot capture nonlinear oscillations and collective bubble interactions, highlighting the need for molecular-scale approaches. Molecular dynamics (MD) simulations inherently resolve interfacial dynamics and nonlinear effects at the atomic scale, making them suitable for exploring such phenomena.

This complexity also poses engineering challenges. To create highly efficient reaction fields, scaling up ultrasonic devices is essential, but cavitation-induced attenuation and nonlinearity become obstacles~\cite{adamou2024}. In addition to bubble interactions, their influence on the acoustic field must be analyzed. Addressing these issues requires extensive parameter searches to identify optimal conditions. While experimental validation is indispensable, numerical simulations are highly promising for condition selection. Since bubble sizes range from microscopic scales at nucleation to macroscopic scales, this is a typical multi-scale, multi-physics problem. Therefore, a direct analysis of phase transition, interfacial dynamics, and acoustic fields at the molecular scale is necessary for a fundamental understanding of ultrasonic cavitation. Considering the strong nonlinearity and non-spherical bubble shapes, a method that naturally handles interfacial dynamics is highly desirable.

Previous approaches to ultrasonic cavitation include experiments, theoretical analyses, and mesoscale simulations~\cite{wu2022,trujillo2018,louisnard2012a,louisnard2012b,commander1989,xu2020}. Experiments mainly use ultrasonic horns to measure bubble dynamics, morphology, and acoustic pressure near the horn. Although direct observations are possible, the resolution of experimental setups limits the ability to capture bubble dynamics over a wide range at high resolution. Recently, X-ray imaging has been employed to analyze bubble cloud formation under ultrasonic irradiation with high resolution~\cite{biasiori2023}. Theoretical analyses based on the Helmholtz equation and mean-field models have enabled qualitative descriptions of acoustic fields but cannot fully account for bubble interactions and nonlinearity~\cite{tiong2025}. With advances in computational power, mesoscale approaches such as Smoothed Particle Hydrodynamics (SPH) and Smoothed Dissipative Particle Dynamics (SDPD) have attracted attention, and hierarchical methods combining SPH particles with microscopic simulators have been recently proposed~\cite{bian2012,litvinov2008,ye2017,morii2021}.

In this study, we aim to directly analyze the early stages of ultrasonic cavitation using large-scale MD simulations. Our goal is to capture the essence of phase transition and interfacial dynamics, which are difficult to address with conventional methods. Phase transition, in particular, is extremely challenging to treat without MD simulations. However, due to computational limitations, previous MD studies have typically been restricted to systems of several hundred million atoms, which can only generate a few cavitation bubbles~\cite{asano2021,asano2022}, making it impossible to reproduce cavitation involving numerous bubbles. To overcome this, we performed MD simulations on the entire Fugaku supercomputer, simulating a system of 100 billion atoms to analyze the initial formation of bubble clouds under ultrasonic irradiation at the molecular level.

This study aims to elucidate the molecular-scale mechanism of the inception of ultrasonic cavitation, which inherently involves phase transition, interfacial formation, and nucleation. These processes dominate the early stage of cavitation and cannot be directly treated by conventional continuum models~\cite{brennen2014, tiong2025} or classical nucleation theory~\cite{blander1975, binder1977}. Therefore, our approach focuses on the molecular dynamics of the inception stage using a large-scale MD simulation, rather than on re-evaluating existing theories.

MD simulations are particularly suitable for resolving the nucleation of bubbles and their early growth stages, which cannot be accessed by experiments or continuum-based simulations. A single-component, impurity-free liquid is deliberately employed to establish a baseline without extrinsic nuclei (e.g., dissolved gases), isolating the intrinsic role of phase transition under ultrasonic driving. This baseline will underpin future extensions to more complex fluids including dissolved gases and surfactants.

Our results show that numerous vapor nuclei appear simultaneously, grow while forming clusters, and that a giant bubble cluster undergoes periodic fragmentation, contraction, and regeneration. These behaviors emerge from molecular-scale phase-transition dynamics and are difficult to capture with existing continuum models. Thus, the present study provides new physical insights into the early-stage cavitation phenomena that are inaccessible to previous theoretical approaches.
\section{Method}
In this study, the fluid was modeled as a monoatomic system, and intermolecular interactions were represented by the smoothed-cutoff Lennard-Jones (sLJ) potential\cite{toxvaerd2011}, given by:
\begin{eqnarray}
  u_{\rm sLJ}(r)&=&
  \begin{cases}
    u_{\rm LJ}(r) - u'_{\rm LJ}(r_{\rm c})(r-r_{\rm c}) - u_{\rm LJ}(r_{\rm c}) & (r \leq r_{\rm c}) \\
    0 & (r > r_{\rm c})
  \end{cases}, \\
  u_{\rm LJ}(r)&=&4\left[ \left(\frac{\sigma}{r}\right)^{12} - \left(\frac{\sigma}{r}\right)^{6} \right],
\end{eqnarray}
where $r$ is the interparticle distance, and $\epsilon$ and $\sigma$ denote the energy and length scales, respectively. The cutoff distance was set to $r_{\rm c}=2.5\sigma$. A prime indicates differentiation with respect to $r$. All physical quantities were expressed in reduced units based on $\epsilon$, $\sigma$, and $\tau = \sqrt{m\sigma^2/\epsilon}$, where $m$ is the particle mass.
The simulation box was a rectangular domain of size $L_x \times L_y \times L_z=5000 \times 6000 \times 6000$. The liquid density was $\rho = 0.6$, and the total number of particles was $N=107~546~400~576$. The acoustic wave propagated along the $x$-direction, with periodic boundary conditions applied in the $y$- and $z$-directions. Vibrating and fixed walls were placed at $x=0$ and $x=5000$, respectively. Each wall was modeled by fixing Lennard-Jones particles on a square lattice with a surface density of $0.80$, resulting in $28~758~528$ particles per wall. The wall particles were not integrated dynamically and acted as an external field on the liquid. A local Langevin thermostat was applied in the region $4500\le x \le 5000$ to absorb acoustic waves~\cite{asano2020,asano2022}. Although a spatially varying friction coefficient would more effectively suppress wave reflections, we employ a constant value of $0.1$ in this study to reduce computational cost.

The initial configuration was generated by randomly placing particles between the walls without overlap, and initial velocities were assigned according to the Maxwell--Boltzmann distribution. The initial state corresponded to the liquid phase at the vapor–liquid coexistence point ($\rho = 0.6, T = 0.85$)~\cite{asano2022}. To equilibrate such a large system, a smaller box of size $5000 \times 250 \times 250$ ($186~712~501$ particles) was first equilibrated and then replicated $24$ times along the $y$- and $z$-directions. Ultrasonic irradiation was applied by oscillating the vibrating wall in the $x$-direction according to:
\begin{eqnarray}
  x_{\rm w}(t)=A\sin\left(2\pi f t\right),
\end{eqnarray}
where the amplitude and frequency were set to $A=15$ and $f=0.005$, respectively. Accordingly, the driving period of the horn was defined as $t_{\rm p} = 200$. The adiabatic sound speed was $c = 2.42$~\cite{asano2022}, and the representative wall velocity was $v_{\rm w} = 2 \pi f A = 0.7$, corresponding to a Mach number of $Ma=0.19$.

Simulations were performed on the Fugaku supercomputer using $152~064$ nodes ($96$\% of the entire system). A hybrid parallelization scheme combining MPI and OpenMP was employed, with $4$ MPI processes per node and $12$ threads per process. Molecular dynamics simulations were conducted using the open-source software LAMMPS (Large-scale Atomic/Molecular Massively Parallel Simulator)~\cite{plimpton1995}. The equations of motion were integrated using the symplectic Verlet algorithm with a time step of $\Delta t = 0.004$. The fastest integration speed achieved was $0.07$ s per step.

To quantify the acoustic field, local quantities are estimated as follows. To this end, the system is divided into subcells with dimensions of $30 \times 30 \times 30$. Local temperature $T_i$ is computed using the equipartition theorem,
\begin{eqnarray}
  T_i = \frac{1}{3n_i}\sum_{j}({\bf v}_j-{\bf v}_{0,i})^2,
\end{eqnarray}
where $i$ denotes the $i$-th subcell, $n_i$ and ${\bf v}_j$ are the number of particles in the subcell and the velocity of the $j$-th particle, respectively.
${\bf v}_{0,i}$ denotes the average velocity of the $i$-th subcell.
The density of $i$-th subcell is computed as
\begin{eqnarray}
  \rho_i=\frac{n_i}{V_i},
\end{eqnarray}
where $V_i$ is the volume of the $i$-th subcell, which is $30\times 30\times 30=27~000$.
In this study, the phase of each subcell was classified based on its density. Since the densities of the vapor and liquid phases at the vapor-liquid coexistence point are $\rho_{\rm vapor}=0.04$ and $\rho_{\rm liq}=0.6$, respectively, the arithmetic average is $0.32$. Subcells with densities lower than $0.32$ were therefore classified as the vapor phase. Vapor-phase subcells connected through any of the $26$ neighboring subcells ($3\times 3\times 3$ connectivity) are grouped into a single vapor cluster.

The pressure is computed using the virial theorem,
\begin{eqnarray}
  p_i=\frac{1}{V_i}\left(\sum_j ({\bf v}_j-{\bf v}_{0,i})^2 + \frac{1}{3}\sum_j {\bf r}_j\cdot {\bf F}_j \right),
\end{eqnarray}
where ${\bf r}_j$ and ${\bf F}_j$ denote the position and the total force acting on the $j$-th particle, respectively.
We also estimated the void fraction to identify the onset of cavitation. The void fraction $\alpha_i$ is defined as
\begin{eqnarray}
  \alpha_{i}=\frac{\rho_{\rm liq}-\rho_i}{\rho_{\rm liq}-\rho_{\rm vapor}}
\end{eqnarray}
where $\rho_{\rm vapor}$ and $\rho_{\rm liq}$ are the densities of vapor and liquid phases at the vapor-liquid coexistence point, and these values are $\rho_{\rm vapor}=0.04$ and $\rho_{\rm liq}=0.6$, respectively.
\section{Results}
\begin{figure*}
  \centering
  \includegraphics[width=\textwidth]{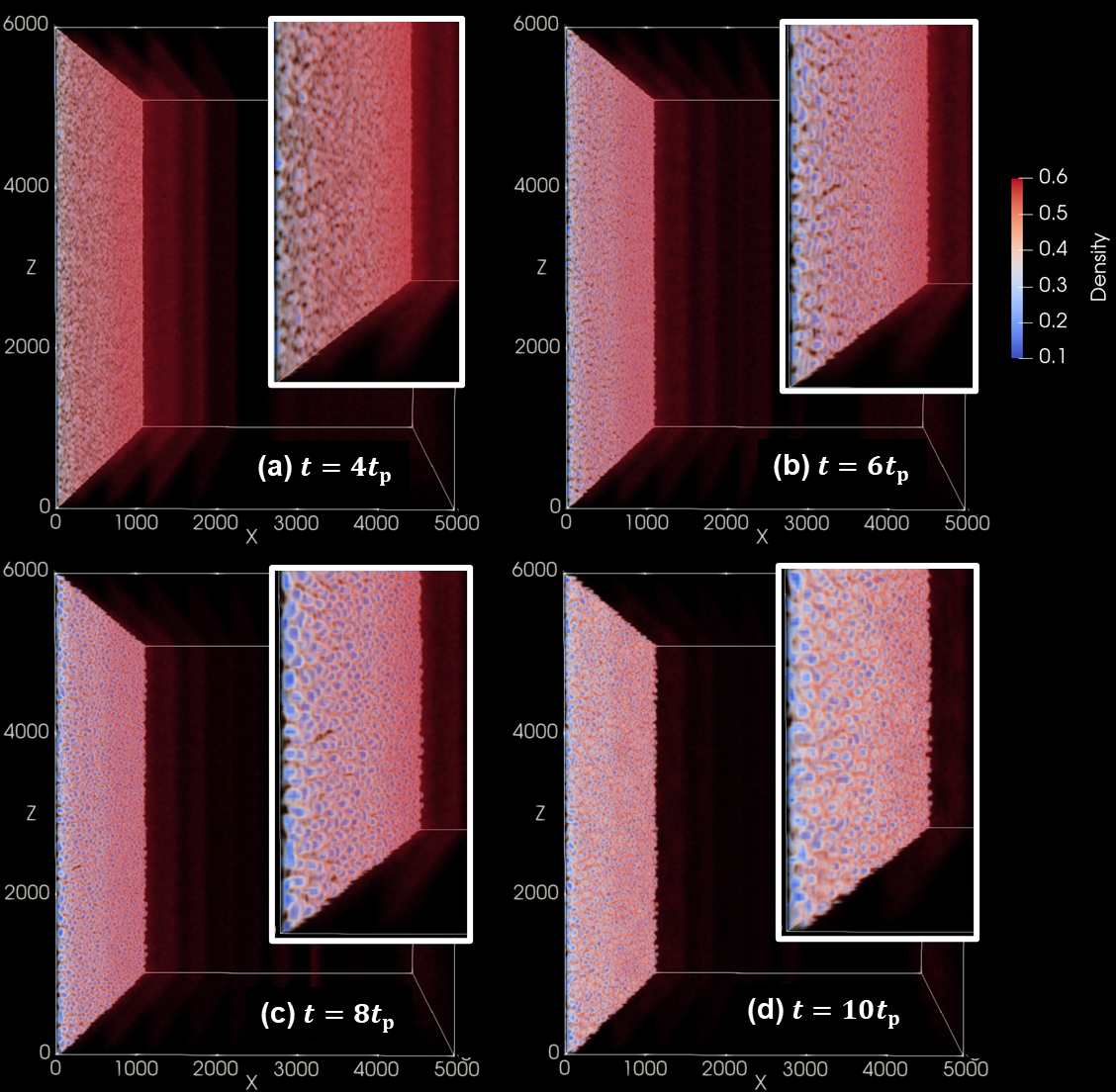}
  \caption{(Color online) Snapshots of the density field at (a) time $t=4t_{\rm p}$, (b) $t=6t_{\rm p}$, (c) $t=8t_{\rm p}$, and (d) $t=10t_{\rm p}$. Time is normalized by $t_{\rm p}$, the driving period of the ultrasonic horn. Only low-density regions are shown to highlight bubble nucleation and growth near the ultrasonic horn. The insets of each panel depict the enlarged view near the horn.}
  \label{fig:1}
\end{figure*}

Figure~\ref{fig:1} shows snapshots of the density fields at $t=4t_{\rm p}, 6t_{\rm p}, 8t_{\rm p}$, and $10t_{\rm p}$. Here, $t_p$ denotes the driving period of the ultrasonic horn, which is used as the time unit for normalization. Each panel of Fig.~\ref{fig:1} includes an enlarged view near the ultrasonic horn. Only the low-density regions are depicted for the sake of visibility. Cavitation bubbles nucleate near the ultrasonic horn and increase in number over time, forming a localized cavitation field similar to experimental observations. The bubble population is almost uniform on the horn surface due to the periodic boundary conditions. It is also found that the bubble shape is highly non-spherical, reflecting strong pressure gradients near the horn. See Supplemental Material for the video corresponding to Fig.~\ref{fig:1}, showing (a) the density field and (b) the enlarged view near the ultrasonic horn~\cite{supplemental}.

\begin{figure*}
  \centering
  \includegraphics[width=\textwidth]{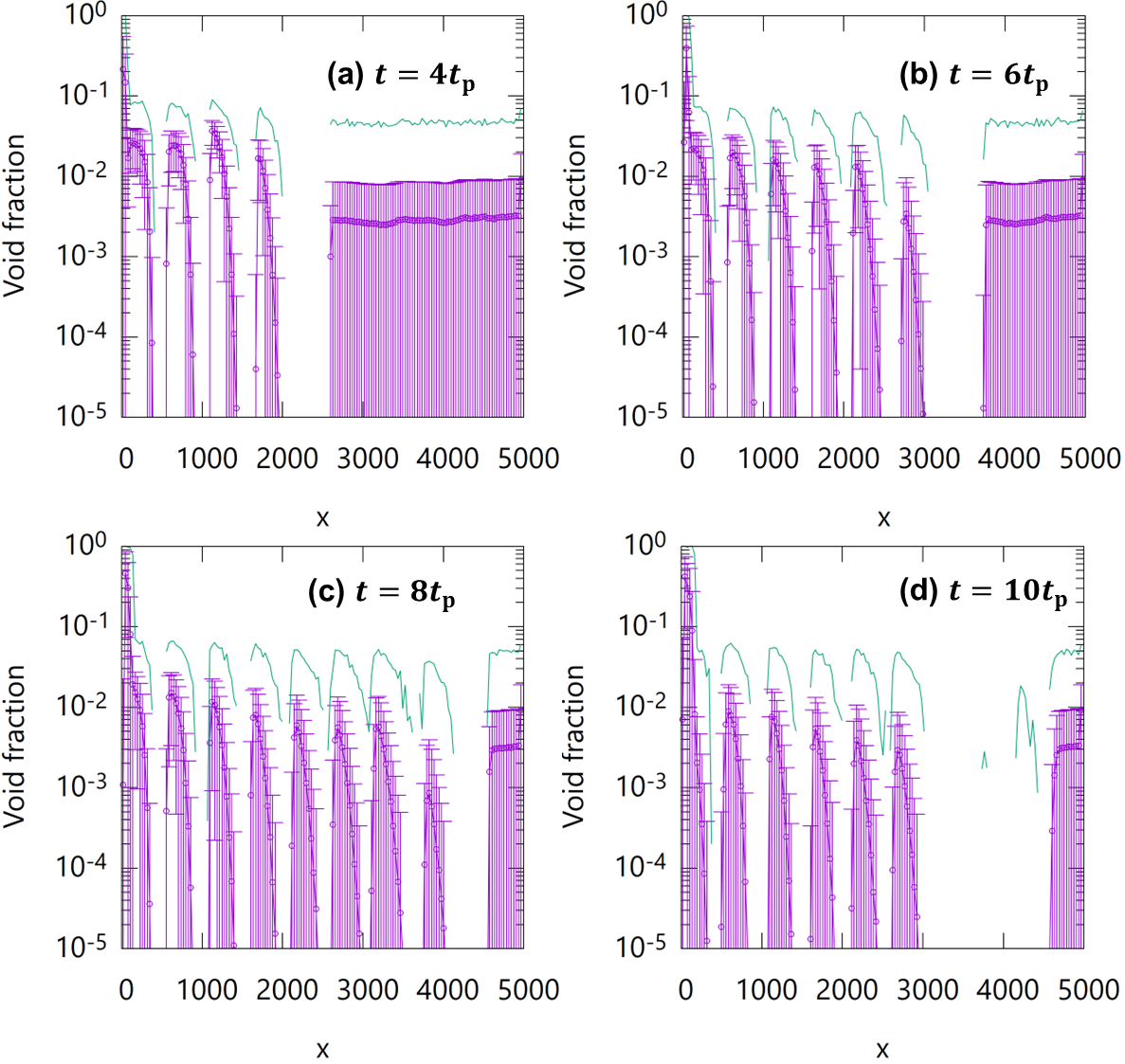}
  \caption{(Color online) Void-fraction distributions at (a) $t=4t_{\rm p}$, (b) $t=6t_{\rm p}$, (c) $t=8t_{\rm p}$, and (d) $t=10t_{\rm p}$. Time is normalized by $t_{\rm p}$. The open-circle symbols connected by a line with error bars represent the mean void fraction and its standard deviation across subcells at each $x$, while the solid curve represents the maximum void fraction. Because the initial state was set at the vapor-liquid coexistence point, a uniform void field ($\sim0.003$) exists at the initial time.}
  \label{fig:2}
\end{figure*}

Figure~\ref{fig:2} shows the void fraction along the $x$-direction at the same time instances. Since the initial condition corresponds to the liquid at the phase transition point, a uniform void fraction of about $0.003$ appears (see right half of Fig.~\ref{fig:2}(a)). Sound-wave propagation causes the inception of the cavitation at the pressure minima. As the pressure wave travels, the void fraction repeatedly increases and decreases, reflecting the periodic creation and collapse of bubbles. This oscillatory behavior is synchronized with the acoustic cycle: void fraction rises during the low-pressure phase and decreases during the high-pressure phase.

Despite these oscillations, the generation of cavitation bubbles remains strongly localized near the horn. In the far field, the void fraction remains much lower than that of near the horn. This indicates that bubble activity does not propagate deeply into the bulk liquid, and cavitation is sustained primarily in the vicinity of the horn.

\begin{figure*}
  \centering
  \includegraphics[width=\textwidth]{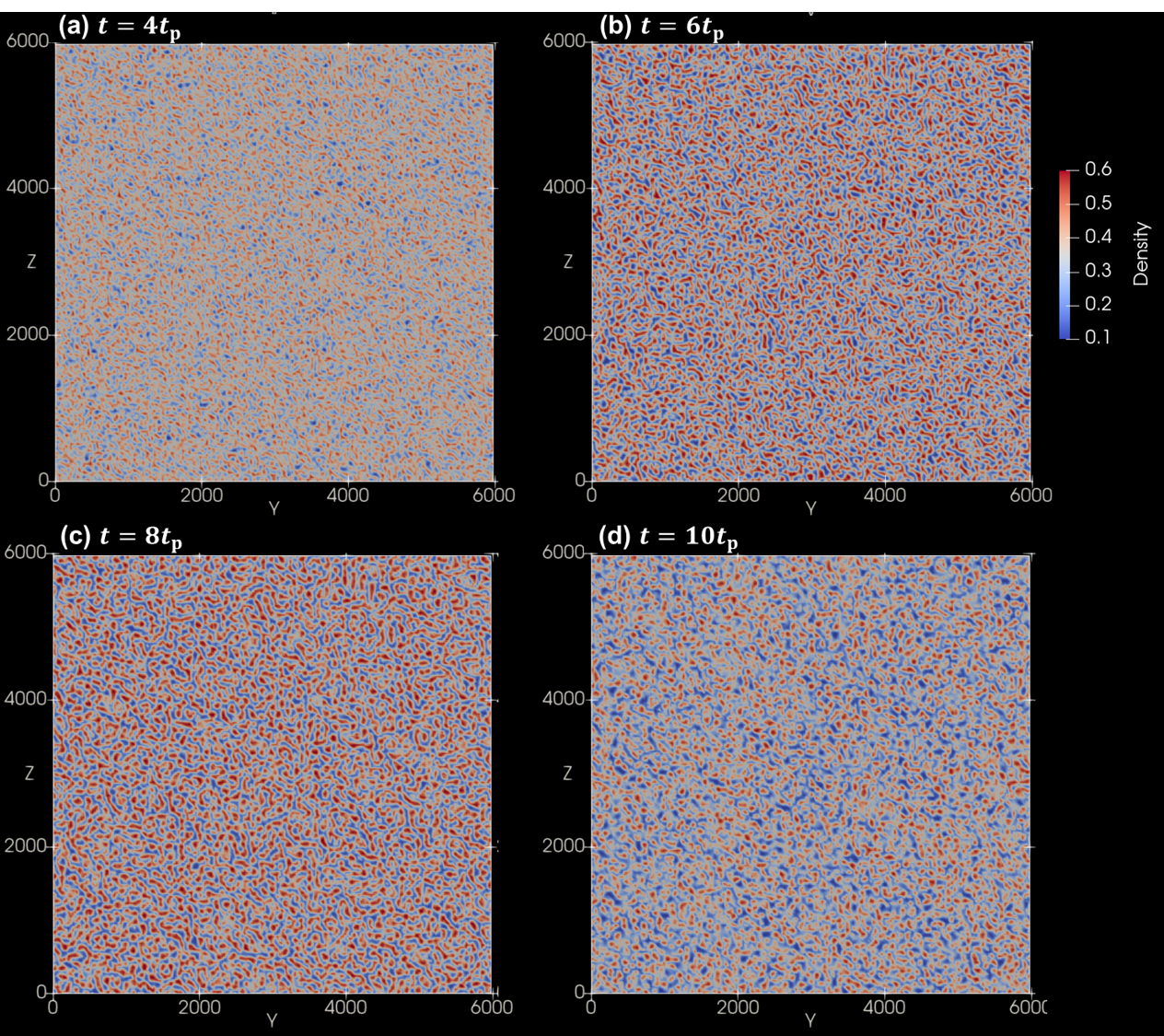}
  \caption{(Color online) Cross-sectional snapshots of the density field on the $yz$-plane at $x=15$ for (a) $t=4t_{\rm p}$, (b) $t=6t_{\rm p}$, (c) $t=8t_{\rm p}$, and (d) $t=10t_{\rm p}$. Time is normalized by $t_{\rm p}$. Numerous small bubbles appear and expand over time.}
  \label{fig:3}
\end{figure*}

Figure~\ref{fig:3} shows cross-sectional views of $yz$-plane at $x=15$ at the same time instances. They show numerous small bubbles uniformly distributed across the plane. The low-density region increases over time and becomes uniformly distributed. The growth of the low-density region implies the clustering of vapor bubbles. See Supplemental Material for the video corresponding to Fig.~\ref{fig:3}, showing (c) the cross-sectional view of the $yz$-plane~\cite{supplemental}.

\begin{figure*}
  \centering
  \includegraphics[width=\textwidth]{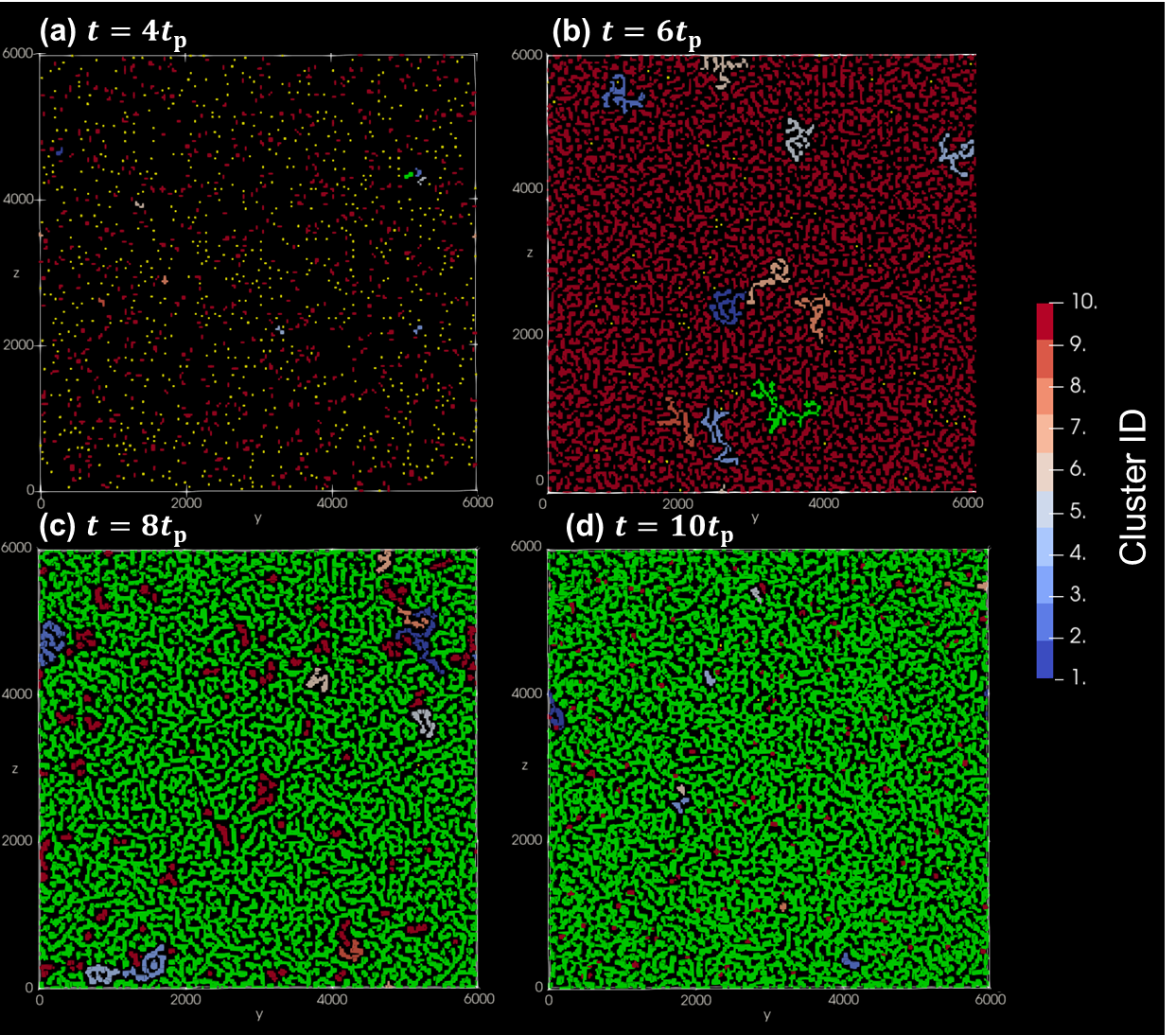}
  \caption{(Color online) Vapor clusters classified by ID at (a) $t=4t_{\rm p}$, (b) $t=6t_{\rm p}$, (c) $t=8t_{\rm p}$, and (d) $t=10t_{\rm p}$. Time is normalized by $t_{\rm p}$. Clusters are labeled in descending order of size, and the largest cluster corresponds to ID = $0$. Single-subcell clusters mainly appear at early times and become negligible as the system evolves. The figure visualizes all vapor-phase clusters in the entire three-dimensional domain, not a cross-sectional slice. The clusters involve the three-dimensional connectivity of bubble structures.}
  \label{fig:4}
\end{figure*}

Figure~\ref{fig:4} shows the cluster map at the same time instances. To estimate the growth of the vapor bubble cluster, the clusters are distinguished by their ID. The ID of the largest cluster is $0$. Single-subcell clusters are prominent at early times (Fig.~\ref{fig:4}(a)) but rapidly diminish as clusters merge into larger connected structures (Fig.~\ref{fig:4}(b)--(d)). The largest cluster grows rapidly over time and eventually covers the horn. It should be noted that part of the horn surface remains uncovered by vapor phase. In previous simulations involving only several hundred million atoms~\cite{asano2022}, the horn surface was entirely covered by a single vapor-phase region, resulting in a simple vapor-liquid coexistence structure. In contrast, the present large-scale system causes a more complex and heterogeneous bubble distribution, with parts of the horn surface remaining uncovered. A similar mesh-like structure has been observed in experiments, and such a pattern is considered a precursor to the cone-shaped bubble clouds that generate extremely intense cavitation field~\cite{moussatov2003}. See Supplemental Material for the video corresponding to Fig.~\ref{fig:4}, showing (d) the growth of bubble clusters~\cite{supplemental}.

\begin{figure*}
  \centering
  \includegraphics[width=\textwidth]{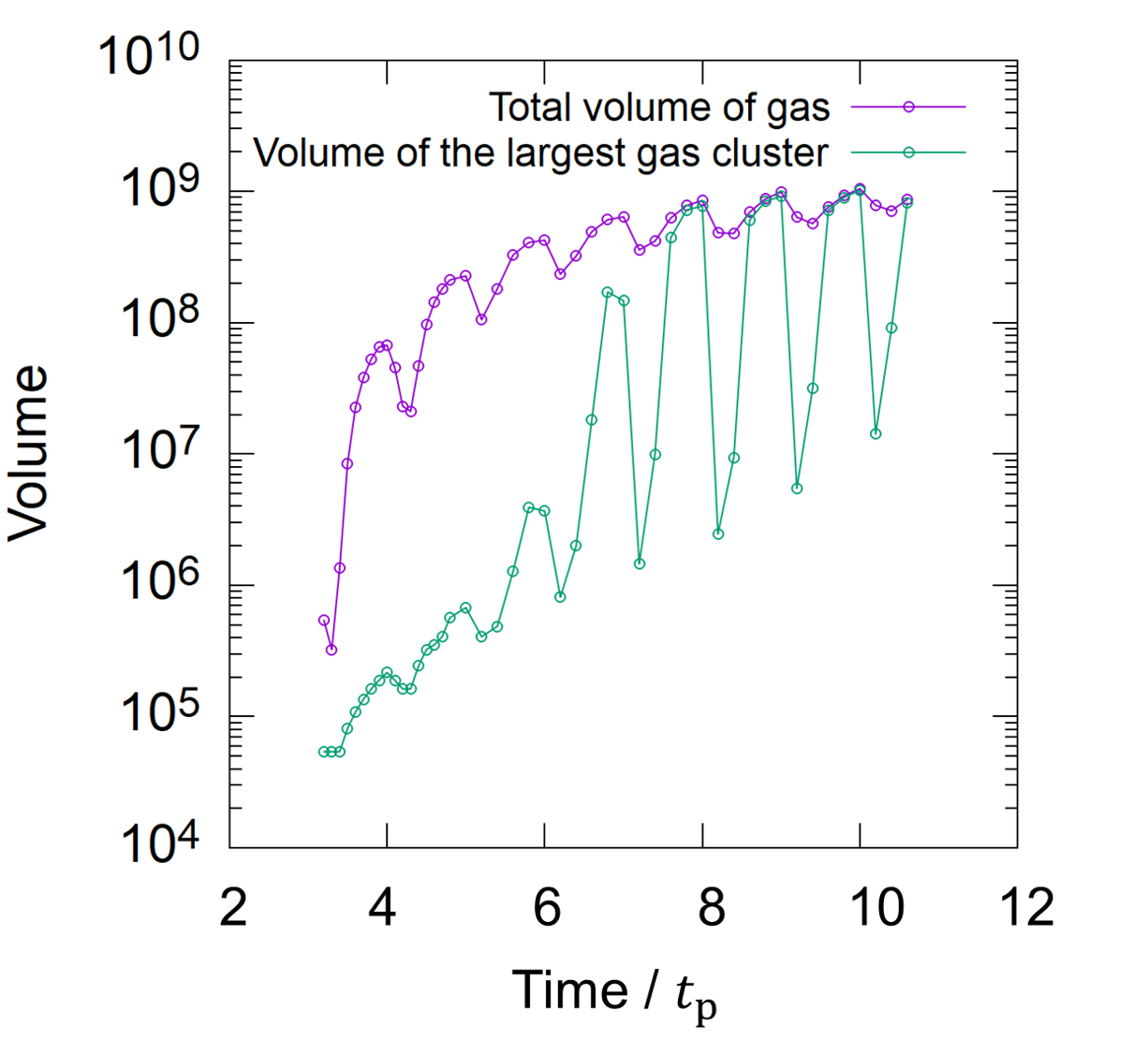}
  \caption{(Color online) Time evolution of the total vapor-phase volume and the volume of the largest vapor cluster. Time is normalized by $t_{\rm p}$. Subcells with $\rho <0.32$ are identified as vapor phase.}
  \label{fig:5}
\end{figure*}

Figure~\ref{fig:5} shows the time evolution of the total volume of vapor region and the volume of the largest vapor cluster. The total volume rapidly increases at $t\simeq 4t_{\rm p}$, then gradually increases with oscillatory behavior. The total volume eventually saturates to a constant value within the range over the simulation time. As for the volume of the largest vapor cluster, it rapidly increases at $t\simeq 7t_{\rm p}$, then oscillates with a large amplitude of several orders of magnitude. The largest cluster repeatedly undergoes fragmentation and recombination with a period corresponding to the driving period of the horn.

\begin{figure*}
  \centering
  \includegraphics[width=\textwidth]{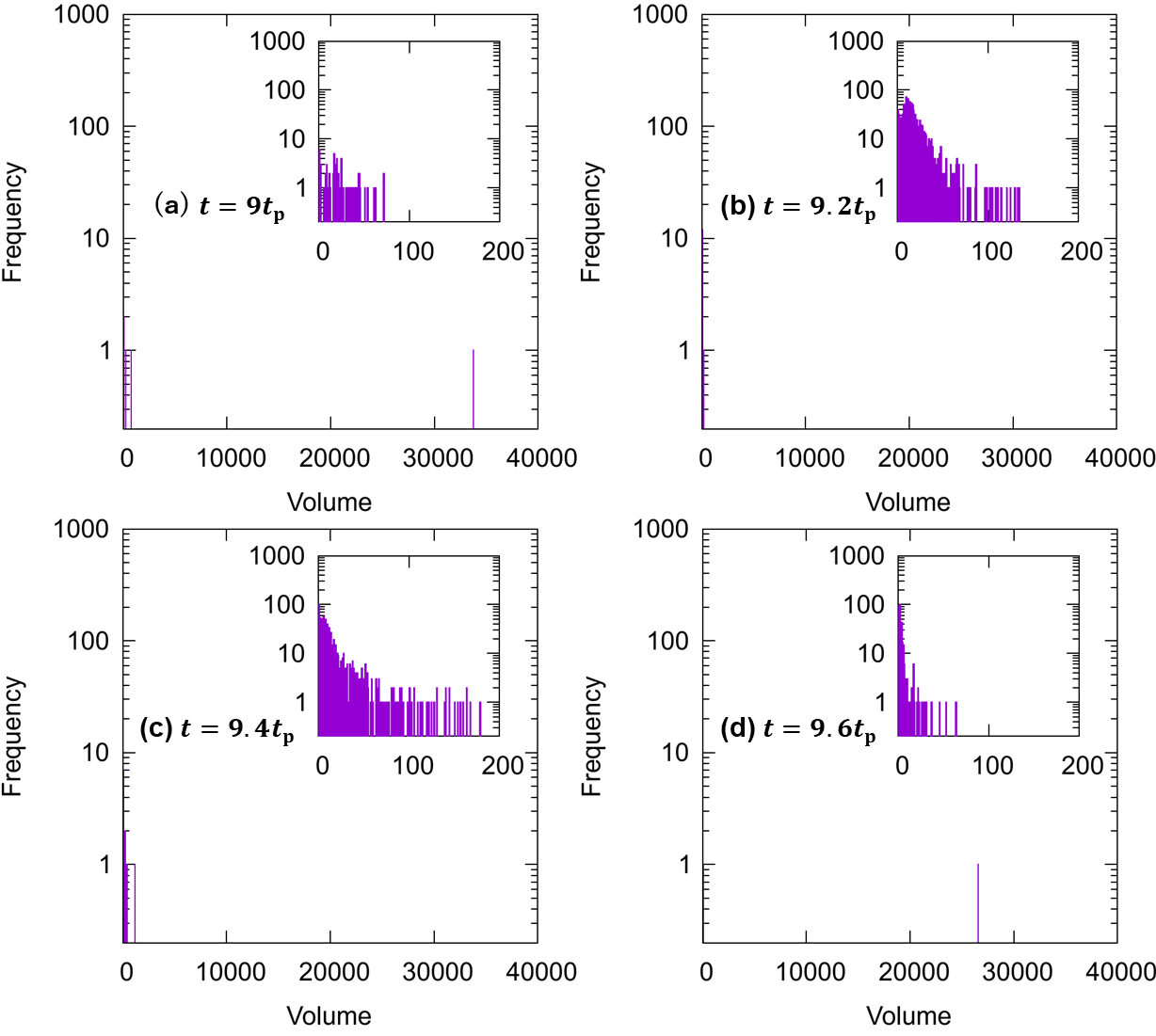}
  \caption{(Color online) Histograms of bubble-cluster sizes at (a) $t=9t_{\rm p}$, (b) $t=9.2t_{\rm p}$, (c) $t=9.4t_{\rm p}$, and (d) $t=9.6t_{\rm p}$, illustrating the periodic fragmentation and recombination of the largest cluster, a process linked to subharmonic-related oscillations. Time is normalized by $t_{\rm p}$.}
  \label{fig:6}
\end{figure*}

In order to investigate how the largest bubble cluster splits and regenerates, the time evolution of the histogram of bubble-cluster sizes was calculated. Figure~\ref{fig:6} shows the histogram of bubble-cluster sizes at $t=9t_{\rm p}$, $t=9.2t_{\rm p}$, $t=9.4t_{\rm p}$, and $t=9.6t_{\rm p}$. At $t=9t_{\rm p}$, the largest cluster, whose volume is about $34~000$, is present, together with a few small clusters with volumes less than $100$ (Fig.~\ref{fig:6}(a)). At this moment, the horn returns to its initial position and subsequently compresses and decompresses the liquid. When the horn compresses the liquid, the largest cluster disappears and several small vapor clusters with volumes of $100$ or less emerge (Fig.~\ref{fig:6}(b)). When the horn decompresses the liquid, these vapor clusters nucleate and grow due to the depressurization (Fig.~\ref{fig:6}(c)). These clusters eventually grow and merge to form a single large vapor cluster at $t=9.6t_{\rm p}$ (Fig.~\ref{fig:6}(d)). The largest vapor cluster then continues to grow and eventually returns to a state similar to Fig.~\ref{fig:6}(a), after which it splits again into the small vapor clusters. This cycle repeats with a period corresponds to the driving period of the horn.

\begin{figure*}
  \centering
  \includegraphics[width=\textwidth]{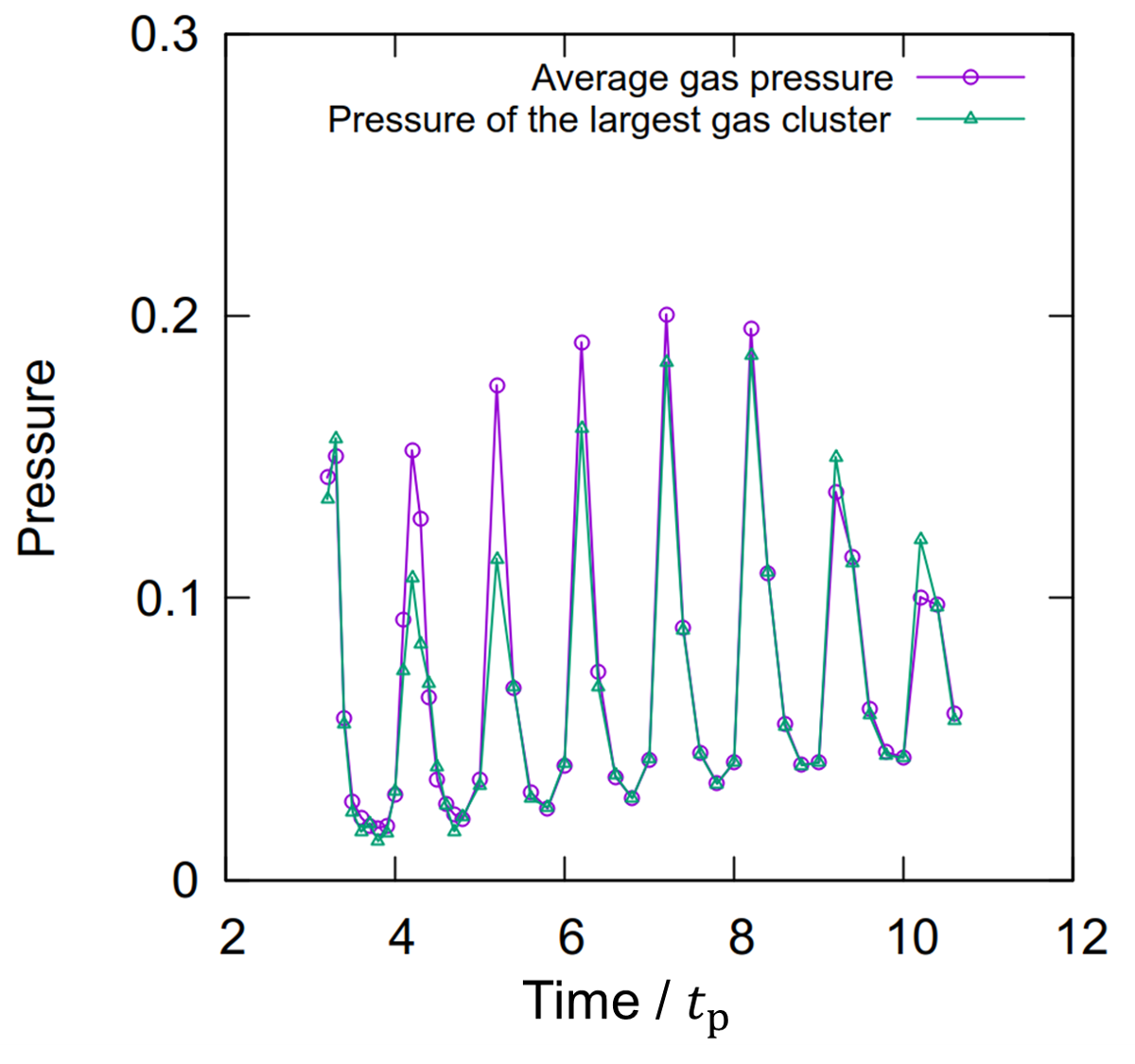}
  \caption{(Color online) Time evolution of the average pressure in the vapor-phase region and within the largest vapor cluster. Time is normalized by $t_{\rm p}$. Pressure was evaluated using the virial theorem; sharp peaks correspond to fragmentation events associated with subharmonic behavior.}
  \label{fig:7}
\end{figure*}

\begin{figure*}
  \centering
  \includegraphics[width=\textwidth]{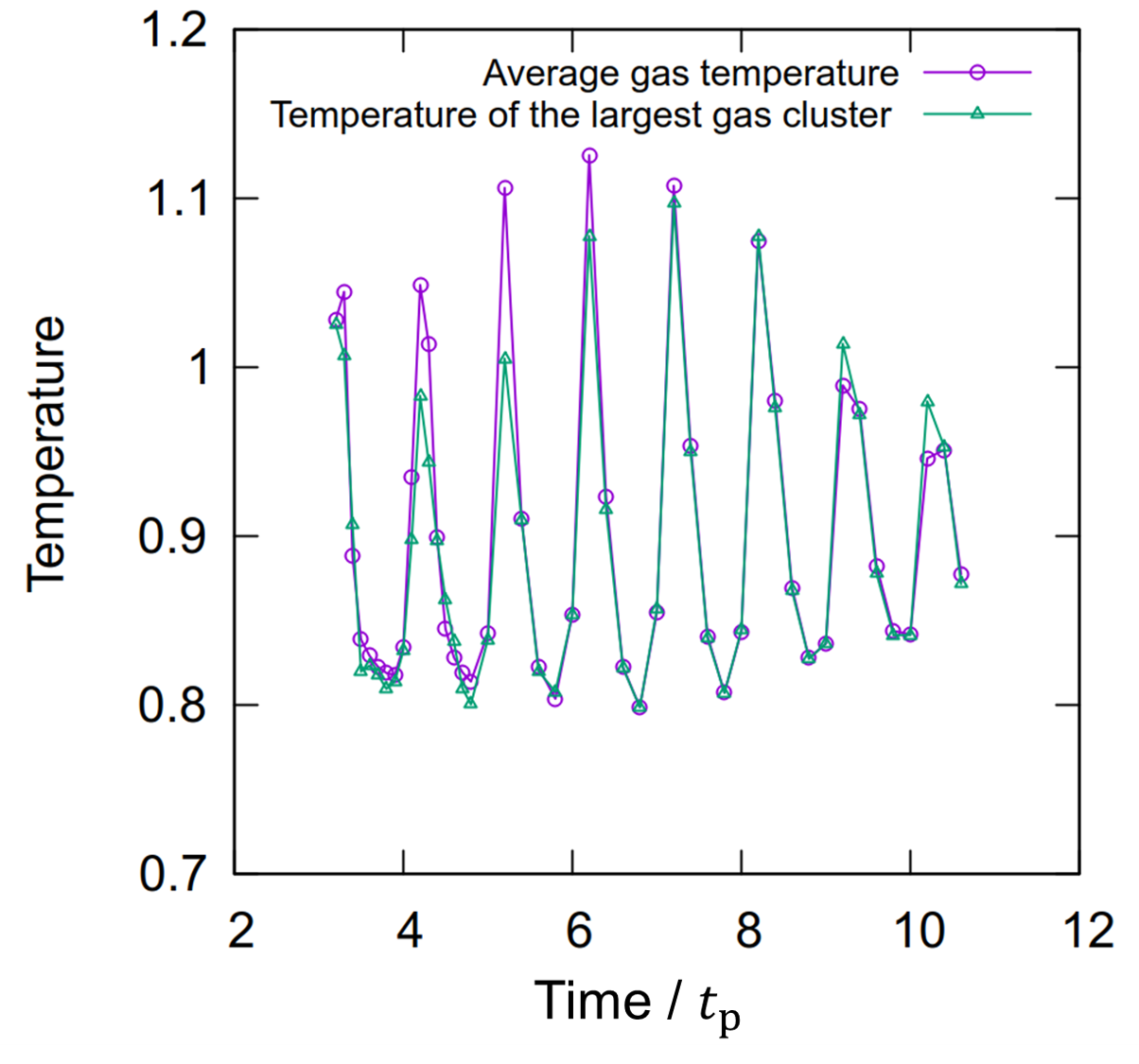}
  \caption{(Color online) Time evolution of the average temperature in the vapor-phase region and within the largest vapor cluster. Temperature was evaluated using the equipartition theorem for kinetic energy.}
  \label{fig:8}
\end{figure*}

Pressure and temperature within bubbles showed oscillatory behavior synchronized with horn vibration, with sharp increases during cluster fragmentation. Figures~\ref{fig:7} and \ref{fig:8} show the time evolution of the pressure and temperature, respectively. Both quantities exhibit the sharp peaks at the moments when the largest cluster undergoes fragmentation. In addition, their amplitudes exhibit oscillatory behavior with a period longer than the driving period of the horn, implying the appearance of subharmonics. This provides a possible mechanism for the experimentally observed subharmonic generation.

\begin{figure*}
  \centering
  \includegraphics[width=\textwidth]{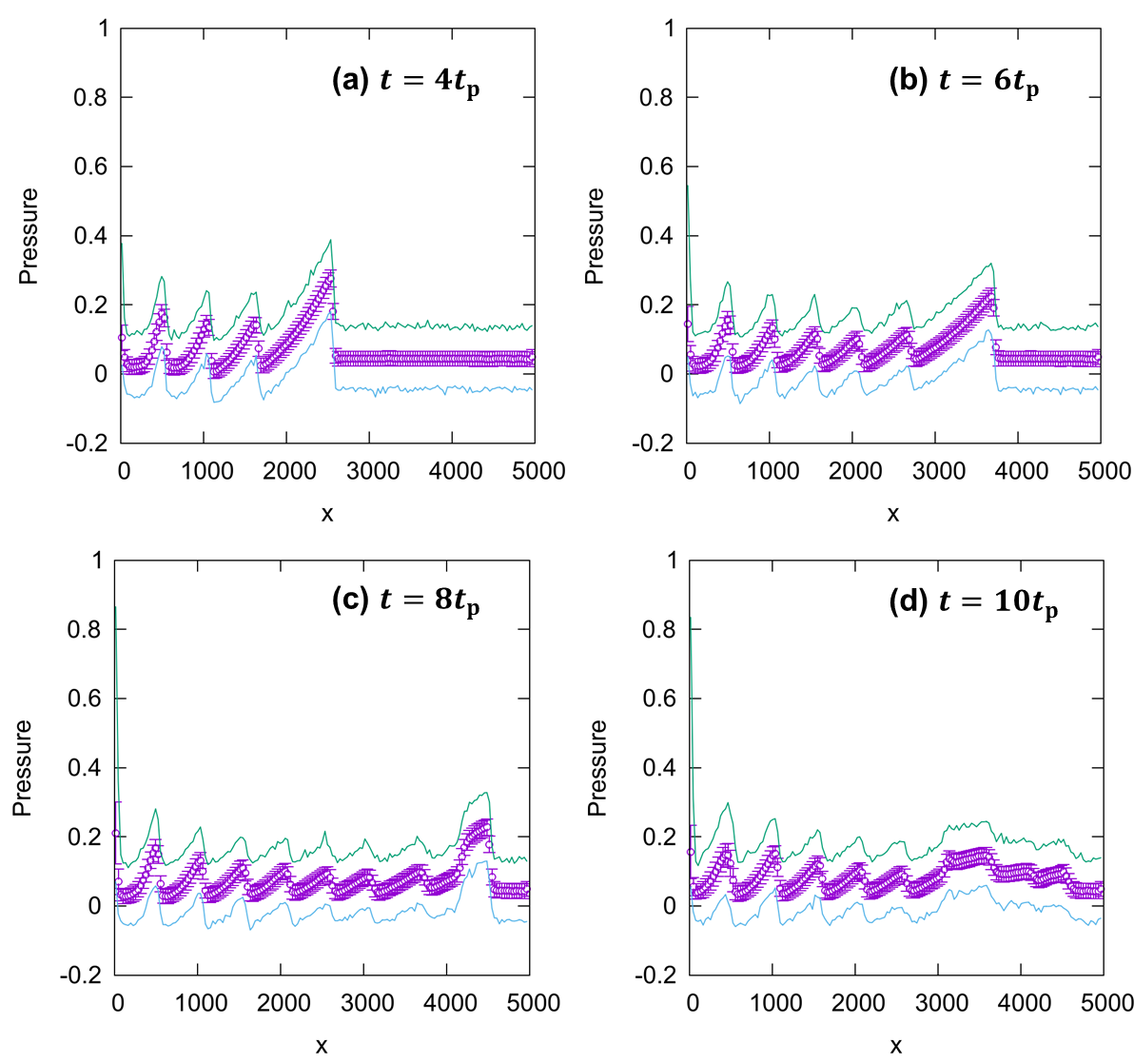}
  \caption{(Color online) Snapshots of pressure waveforms at (a) $t=4t_{\rm p}$, (b) $t=6t_{\rm p}$, (c) $t=8t_{\rm p}$, and (d) $t=10t_{\rm p}$. Time is normalized by $t_{\rm p}$. The open-circle symbols connected by a line with error bars indicate the mean pressure and its standard deviation across subcells at each $x$, while the solid curves represent the maximum and minimum pressures at each $x$ position, respectively.}
  \label{fig:9}
\end{figure*}

Figure~\ref{fig:9} shows the spatial distribution of the sound field. Sound-waves generated from the horn propagate while undergoing attenuation, with the maximum pressure consistently occurring in the vicinity of the horn. In contrast, although the negative pressure also appears in the far-field, actual cavitation is confined to the region immediately adjacent to the horn. This is because bubble formation is governed not simply by the absolute minimum pressure value, but by the intense pressure gradients and high-speed interfacial motion near the horn. In Fig.~\ref{fig:9}(d), it can also be observed that the reflected waves from the fixed boundary cause local fluctuations in the pressure amplitude.

\begin{figure*}
  \centering
  \includegraphics[width=\textwidth]{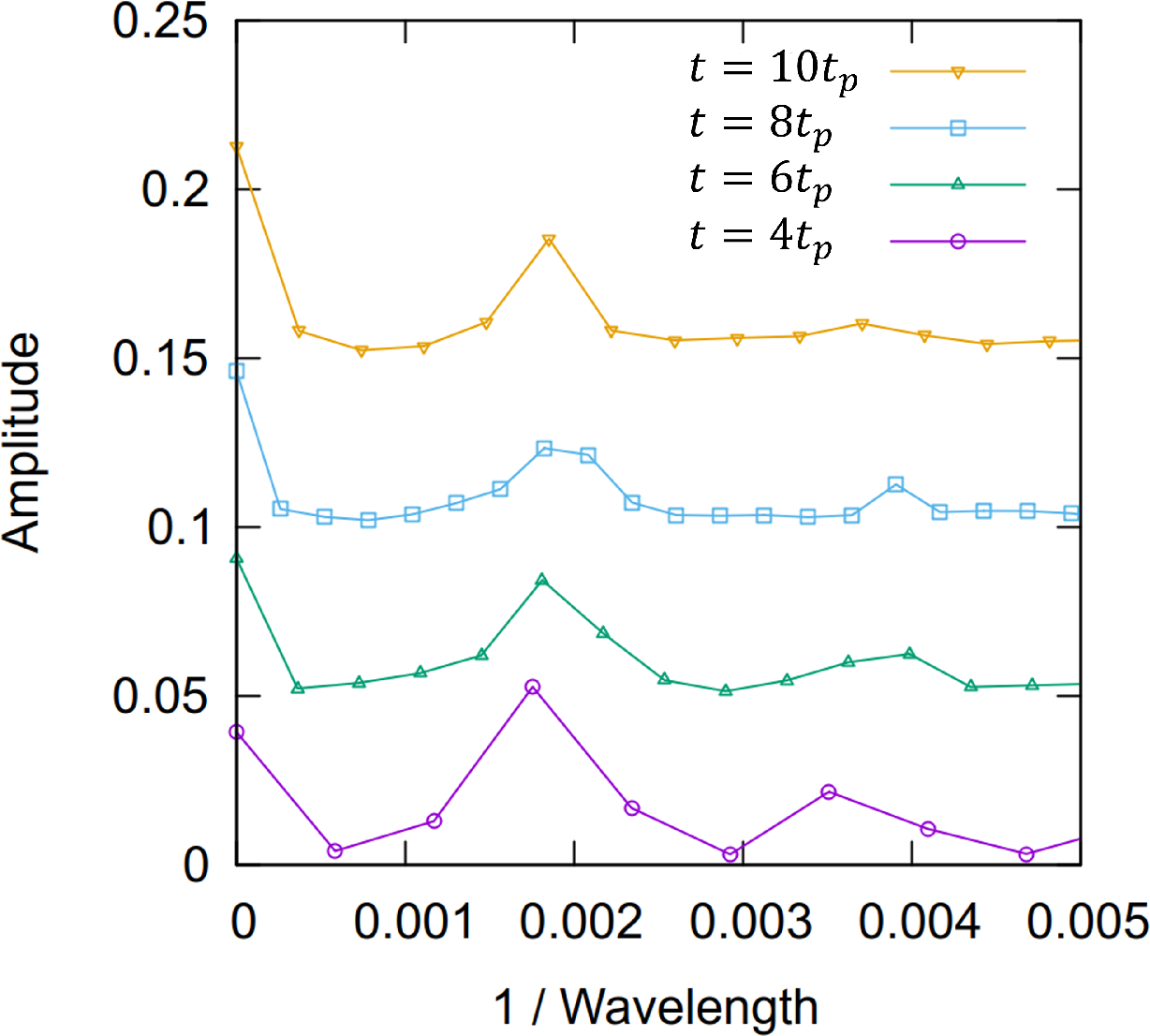}
  \caption{(Color online) Power spectra of the pressure waveforms in Fig.~\ref{fig:9} obtained by Fourier analysis. The spectra are vertically offset for clarity.}
  \label{fig:10}
\end{figure*}

Despite the formation of bubbles, the effect on sound-wave propagation was minimal, as bubbles remained confined near the horn surface. Figure~\ref{fig:10} shows the Fourier transform of the waveforms in Fig.~\ref{fig:9}. To focus on the intrinsic acoustic properties of the medium, the Fourier analysis was performed on the stable portion of the waveforms, excluding the initial nonlinear wave-front and the subsequent reflected waves from the boundary. The peak positions of the inverse wavelength are located at $\lambda^{-1}\simeq0.0019$, corresponding to a sound speed of about $2.6$. Although this value slightly deviates from the previously reported $2.42$~\cite{asano2022}, it remains reasonably consistent within the numerical resolution $\Delta \lambda^{-1}\simeq 0.0002$ determined by the system size $L_x$.

In addition, no abrupt decrease in the amplitude, which corresponds to the abrupt increase in the attenuation coefficient, was observed near the horn, suggesting that cavitation does not induce rapid damping in its vicinity. Therefore, no significant influence on the sound speed or attenuation was detected, supporting the conclusion that cavitation localized at the horn has limited influence on bulk acoustic properties. These results indicate that bubble growth requires not only low-pressure regions but also the horn-induced interfacial acceleration, underscoring its critical role in cavitation dynamics.

%These results demonstrate that bubble dynamics are strongly coupled with horn oscillation and suggest a mechanism for subharmonic generation.
%

\section{Summary and Discussion}
This study demonstrates, for the first time, the reproduction of multi-bubble cavitation dynamics at the molecular scale using an unprecedented 100-billion-atom MD simulation. We elucidated the initial formation of bubble clouds and uncovered a striking phenomenon: giant bubble clusters undergo periodic collapse and regeneration in strict synchrony with horn oscillation. Sharp pressure and temperature spikes during fragmentation suggest a strong connection to subharmonic generation, a hallmark of sonochemical processes.

  The critical size of the vapor nuclei is estimated based on the classical nucleation theory~\cite{blander1975} as $R_{\rm c}=2\gamma / \Delta p=\mathcal{O}(10)$, where $\gamma=0.32$~\cite{watanabe2012} and $\Delta p=0.1$ (from Fig.~\ref{fig:9}). Since the bubble equivalent radius $R$ derived from the vapor volume assuming a spherical shape observed in the MD simulations ranges from $20$ to $600$, these values are consistent with the classical nucleation theory. The oscillation behavior of the bubbles is also consistent with the natural frequency ignoring surface tension~\cite{minnaert1933}:
  \begin{eqnarray}
    f_{\rm n} = \frac{1}{2\pi R} \left(\frac{3\kappa p}{\rho}\right)^{\frac{1}{2}},
  \end{eqnarray}
where $\kappa$ is the ratio of specific heat, for which the ideal gas value of $5/3$ was used. The resonance frequency corresponding to the simulated bubble sizes lies in the range $0.0002\le f \le0.007$. Since the driving frequency of the horn is $f=0.005$, the expansion–contraction cycle of the bubbles is expected to synchronize with the horn oscillation.  The time evolution shown in Fig.~\ref{fig:5} indicates that the observed bubble-cluster dynamics are physically consistent with classical resonance theory~\cite{rayleigh1917}.

Our analysis reveals that acoustic propagation is minimally affected by cavitation, highlighting attenuation control as a critical factor for scaling ultrasonic systems. Moreover, horn kinematics emerge as a decisive driver of cavitation intensity and spatial distribution. The observed synchronization between horn motion and cluster fragmentation underscores the need for precise control of oscillatory dynamics. Long-period oscillations and nonlinear bubble interactions further point to collective effects as key contributors to cavitation behavior, consistent with experimental observations. These findings provide molecular-scale evidence of mechanisms underlying sonochemistry and offer practical guidance for designing efficient ultrasonic reactors and biomedical technologies.

We also found that bubble growth coincides with the arrival of reflected acoustic waves at the container wall, indicating that global geometry strongly influences cavitation dynamics. Larger domains may alter synchronization between wave reflection and cluster fragmentation, potentially modifying subharmonic behavior. However, MD simulations face strict size limitations due to computational cost, making boundary conditions that suppress reflection essential for accurate modeling without excessive overhead. This insight is critical for extending MD approaches to realistic ultrasonic systems.

Future work will focus on quantifying long-period oscillations, analyzing attenuation in larger domains, and exploring variations in driving frequency and amplitude. Extending simulations to complex fluids and surfactant-containing systems will bridge the gap to real-world applications such as drug delivery and non-invasive therapies. A central challenge will be clarifying the origin of subharmonic generation through systematic variation of horn geometry and amplitude, as this phenomenon is pivotal for enhancing sonochemical efficiency.
\begin{acknowledgment}
  The author would like to thank H. Noguchi for fruitful discussions. The author also expresses sincere gratitude to Y. Ota, E. Tomiyama, and N. Sueyasu of the Research Organization for Information Science and Technology (RIST) for their extensive assistance with the use of the supercomputer Fugaku and the LAMMPS environment on Fugaku. This work used computational resources of the supercomputer Fugaku provided by the RIKEN Center for Computational Science (Project ID: hp240112).This research was partially supported by JSPS KAKENHI, Grant No. JP23K03242. The author also acknowledges the Supercomputer Center, Institute for Solid State Physics (ISSP), the University of Tokyo, for allowing the author to use their supercomputing facilities.
%\acknowledgment
%For environments for acknowledgement(s) are available: \verb|acknowledgment|, \verb|acknowledgments|, \verb|acknowledgement|, and \verb|acknowledgements|.
\end{acknowledgment}

%\appendix
%\section{}

%Use the \verb|\appendix| command if you need an appendix(es). The \verb|\section| command should follow even though there is no title for the appendix (see above in the source of this file).

%For authors of Invited Review Papers, the \verb|profile| command is prepared for the author(s)' profile.  A simple example is shown below.

%\begin{verbatim}
%\profile{Taro Butsuri}{was born in Tokyo, Japan in 1965. ...}
%\end{verbatim}

\bibliographystyle{jpsj}  % 他にも jplain, unsrt, ieeetr などあり

%\bibliography{references}  % references.bib というファイル名の場合

%\begin{thebibliography}{9}
%\bibitem{jpsj} The abbreviation for JPSJ must be ``J. Phys. Soc. Jpn." \note{in the reference list}.
%\bibitem{instructions} More abbreviations of journal titles are listed in ``Instructions for Preparation of Manuscript".
%\bibitem{etal} The use of ``et al.'' is not accepted in principle, therefore, all the authors must be listed.
%\bibitem{ibid} The term ``ibid.'' should not be used even if the same journal or book is cited with different page numbers.
%\bibitem{Errata} Errata should be listed under the same reference number. 
%\end{thebibliography}

\end{document}